\documentclass[epjST]{svjour}
\usepackage{graphics}
\usepackage{amsmath}

 \newcommand{\ave}[1]{{\langle{#1}\rangle}}

\begin{document}
\title{Inhomogeneous phases in the quark-meson model with explicit chiral-symmetry breaking}
\author{Michael Buballa\inst{1}\fnmsep\thanks{\email{michael.buballa@tu-darmstadt.de}} \and Stefano Carignano \inst{2} \and Lennart Kurth\inst{1} }
\institute{Technische Universit\"at Darmstadt, Department of Physics, Institut f\"ur Kernphysik, Theoriezentrum 
\and Departament de F\'isica Qu\`antica i Astrof\'isica and Institut de Ci\`encies del Cosmos, Universitat de Barcelona}
\abstract{
We investigate the existence of
inhomogeneous chiral phases in the quark-meson model with explicit chiral-symmetry breaking. 
We find that the inhomogeneous region shrinks with increasing pion masses but survives for the physical value of $m_\pi$.
The instability towards inhomogeneous matter occurs in the scalar channel, while pseudoscalar modes are disfavored.
} 
\maketitle

\section{Introduction}
\label{sec:intro}

Mapping the phase diagram of QCD at nonvanishing temperature $T$ and quark chemical potential $\mu$ is one of the 
major challenges in strong-interaction physics \cite{Kumar:2013cqa,Friman:2011zz}.
Lattice QCD calculations at $\mu = 0$ revealed that chiral symmetry, which is spontaneously broken in vacuum, gets approximately restored in a smooth crossover at $T \sim 150$~MeV \cite{Aoki:2006we}, while in the regime of low $T$ and nonzero $\mu$, where standard lattice methods are not applicable, model studies as well as continuum approaches to QCD indicate the possibility of a first-order phase transition, terminating at a second-order critical endpoint (CEP)  \cite{Asakawa:1989bq,Scavenius:2000qd,Fischer:2019,Fu:2019hdw}. Most of these calculations rely however on the assumption that these phases are homogeneous, i.e., that the chiral order parameter does not vary in space. 
Allowing for spatially non-uniform order parameters, such inhomogeneous phases often turn out to be favored in some
region of the phase diagram, typically covering parts of or even the entire first-order boundary between the homogeneous phases.
Specifically this was found within the Nambu--Jona-Lasinio (NJL) \cite{Nakano:2004cd,Nickel:2009wj} and the Quark-Meson (QM) model \cite{Nickel:2009wj,Carignano:2014jla}, but also in QCD using Dyson-Schwinger equations \cite{Muller:2013tya}. (For a review about inhomogeneous chiral phases, see Ref.~\cite{Buballa:2014tba}.) 

However, most of these studies have been performed in the chiral limit, while the situation for the more realistic case with a small explicit breaking of chiral symmetry is less clear. For the NJL model it was found that the inhomogeneous phase shrinks when a 
nonvanishing bare quark mass is introduced but is still present for realistic masses \cite{Nickel:2009wj}. More generally it was shown in Ref.~\cite{Buballa:2018hux} that the inhomogeneous phase always reaches up to the CEP in that model and thus survives as long as there is a first-order phase transition in the homogeneous case. For the QM model, on the other hand, it was found in Ref.~\cite{Andersen:2018osr} that the inhomogeneous phase becomes disfavored already for a rather small amount of explicit symmetry breaking, corresponding to a pion mass of about one quarter of the physical value. The calculation was however done only for one specific spatial modulation of the order parameter, a so-called chiral density wave (CDW). This modulation is relatively simple to handle but is known not to be the most favored shape in most cases, and away from the chiral limit it is not even a self-consistent solution.

In the present work we therefore study the effect of explicit chiral-symmetry breaking on inhomogeneous phases in the QM model, starting with a stability analysis of the homogeneous phase. This method, which has already been employed in Ref.~\cite{Buballa:2018hux} to the analogous problem in the NJL model, has the advantage that one does not need to know the explicit shape of the spatial modulation. It only relies on the assumption that at the phase boundary the homogeneous phase becomes unstable against small inhomogeneous fluctuations, i.e., that the phase transition is of second order. This analysis will therefore yield a sufficient criterion for the inhomogeneous regime in the model, while the true inhomogeneous phase can be larger.
We will then support the results of the stability analysis with a calculation of the full model phase diagram employing a specific self-consistent ansatz away from the chiral limit, as well as by a Ginzburg-Landau expansion close to the CEP. 

The remainder of this article is organized as follows:
We introduce our theoretical framework in Sec.~\ref{sec:theory}, then in Sec.~\ref{sec:parameters} we discuss the QM parameter fitting procedure away from the chiral limit. We show our numerical 
results for the phase diagram in Sec.~\ref{sec:phase} and discuss our conclusions in Sec.~\ref{sec:conclusions}.

\section{Theoretical framework}
\label{sec:theory}

We consider the QM model defined by the Lagrangian
\begin{equation}
\mathcal{L}_\mathrm{QM}
=
\bar{\psi}
\left(
i\gamma^\mu \partial_\mu
-
g(\sigma+i\gamma_5 \mbox{\boldmath$\tau$} \cdot \mbox{\boldmath$\pi$})
\right)
\psi
+
\mathcal{L}_\mathrm{M}^\mathrm{kin}
-
U(\sigma,\mbox{\boldmath$\pi$})
\ ,
\end{equation}
where $\psi$ is a quark spinor field with $N_f = 2$ flavor and $N_c = 3$ color degrees of freedom, 
coupled via a Yukawa interaction with coupling constant $g$ to the scalar sigma meson $\sigma$ 
and the pseudoscalar pion triplet  \mbox{\boldmath$\pi$}. Here \mbox{\boldmath$\tau$}$=(\tau_1,\tau_2,\tau_3)$ denotes the 
Pauli matrices in isospin space.
The meson kinetic contributions read
\begin{equation}
\mathcal{L}_\mathrm{M}^\mathrm{kin} = \frac{1}{2} \left(
\partial_\mu\sigma \partial^\mu\sigma 
+
\partial_\mu \mbox{\boldmath$\pi$} \cdot \partial^\mu \mbox{\boldmath$\pi$}
\right) \ ,
\end{equation}
and
\begin{equation}
\label{eq:Uqm}
U(\sigma,\mbox{\boldmath$\pi$})
=
\frac{\lambda}{4}
\left(
\sigma^2
+
 \mbox{\boldmath$\pi$}^2
-
v^2
\right)^{2}
-
c\sigma
\,
\end{equation}
is the meson potential.  
In the limit $c=0$ it is symmetric unter $O(4)$ transformations of the meson vector $\phi = (\sigma,  \mbox{\boldmath$\pi$})$,
which can be identified with the chiral $SU(2)_L\times SU(2)_R$ symmetry. 
For $c\neq 0$ the $O(4)$ symmetry is broken explicitly down to $O(3)$, corresponding to the $SU(2)$ isospin symmetry.
The model parameters, $g$, $\lambda$, $v^2$, and $c$, will be fitted to vacuum properties as discussed in Sec.~\ref{sec:parameters}.

The thermodynamic properties of the model are encoded in the grand potential per volume $V$, $\Omega(T,\mu) = -\frac{T}{V} \log\mathcal{Z}(T,\mu)$, where $\mathcal{Z}(T,\mu)$ denotes the grand canonical partition function, which depends on the temperature $T$ and the quark chemical potential $\mu$. 
In the following we perform the mean-field approximation, replacing the quantum fields $\sigma$ and  \mbox{\boldmath$\pi$}
by  their expectation values, i.e., by classical fields. 
We assume that these fields are time independent but we retain their dependence on the spatial coordinate  \mbox{\boldmath$x$} in order to allow for inhomogeneous phases. 
Moreover, we assume that only the third isospin component of the pion field develops a nonvanishing expectation value, which we call $\pi(\mbox{\boldmath$x$})$, while, for simplicity, we keep the name $\sigma(\mbox{\boldmath$x$})$ for the classical sigma field.
The mean-field grand potential per volume (``thermodynamic potential'') is then given by
\begin{equation}
\Omega_\mathrm{MFA}(T,\mu;\sigma, \pi)
=
\Omega_\mathrm{q}(T,\mu;\sigma, \pi) + \Omega_\mathrm{mes}(\sigma, \pi) \,, 
\label{eq:OmegaMFA}
\end{equation}
with a purely mesonic part
\begin{equation}
\Omega_\mathrm{mes} 
=
\frac{1}{V}\int_V d^3x\,\left\{
       \frac{1}{2}\left( (\mbox{\boldmath$\nabla$}\sigma(\mbox{\boldmath$x$}))^2  + (\mbox{\boldmath$\nabla$}
         \pi(\mbox{\boldmath$x$}))^2 \right) 
       +  U(\sigma(\mbox{\boldmath$x$}), \pi(\mbox{\boldmath$x$}))
\right\}
\end{equation}
and a quark part
\begin{equation}
\Omega_\mathrm{q} = 
-\frac{T}{V}
\mathbf{Tr} \, \mathrm{log} \frac{S^{-1}}{T},
\label{eq:Omegaq}
\end{equation}
where 
\begin{equation}
S^{-1}(x) 
=
 i\gamma^\mu \partial_\mu + \mu\gamma^0
-
g(\sigma(\mbox{\boldmath$x$})+i\gamma_5 \tau_3 \pi(\mbox{\boldmath$x$}))
\label{eq:Sinv}
\end{equation}
corresponds to the inverse dressed quark propagator at chemical potential $\mu$
in the presence of the sigma and pion mean fields, and
$\mathbf{Tr}$ denotes a functional trace running over the Euclidean four volume $V_4 = [0,\frac{1}{T}] \times
V$ as well as color, flavor and spinor degrees of freedom.
These expressions are basically identical to those in Ref.~\cite{Carignano:2014jla},
where the same model in the chiral limit was considered. The only exception is that we now have to take into account
the explicitly symmetry-breaking term $-c\sigma$ in the mesonic potential.

\subsection{Stability analysis}
\label{sec:stability}

In order to determine the ground state of the system at given $T$ and $\mu$, we must minimize the thermodynamic potential with respect to the mesonic fields $\sigma$ and $\pi$. 
While this is standard for spatially constant mean fields, the functional minimization of $\Omega_\mathrm{MFA}$ 
with respect to arbitrary non-uniform fields is obviously a much harder problem, which has not yet been solved in full glory for
$3+1$ space-time dimensions. Instead of tackling the full problem, one possibility is to
 perform a stability analysis, applying the same methods, which have been used in Ref.~\cite{Buballa:2018hux}  to analyze inhomogeneous phases in the NJL model.
To this end we split the meson fields into spatially constant parts, corresponding to the lowest homogeneous state of the system, and small fluctuations with arbitrary spatial shapes. 
Since in homogeneous systems the pion field is disfavored against the sigma field due to the symmetry-breaking term in the potential,
the constant part appears only in the sigma sector, i.e., we have 
\begin{equation}
\sigma(\mbox{\boldmath$x$}) = \bar{\sigma} + \delta\sigma(\mbox{\boldmath$x$}), \quad
\pi(\mbox{\boldmath$x$}) = \delta\pi(\mbox{\boldmath$x$}),
\end{equation}
where $\bar{\sigma}$ corresponds to the (in general $T$ and $\mu$ dependent) value of the sigma field in the homogenous
ground state, and $\delta\sigma$ and $\delta\pi$ are the fluctuations.

Plugging this into Eqs.~(\ref{eq:OmegaMFA}) -- (\ref{eq:Sinv}), the thermodynamic potential can be decomposed as
\begin{equation}
\Omega_\mathrm{MFA}(T,\mu;\sigma, \pi)
=
\sum\limits_{n=0}^\infty \Omega^{(n)},
\end{equation}
with $\Omega^{(n)}$ being of the $n$th order in the fluctuations.
Specifically one finds for the contributions up to quadratic order
\begin{eqnarray}
\Omega^{(0)} &=& 
-\frac{T}{V}
\mathbf{Tr} \, \mathrm{log} \frac{S_0^{-1}}{T} + U(\bar{\sigma},0),
\\[1mm]
\Omega^{(1)} &=& \frac{T}{V}
\mathbf{Tr} \, \left( S_0 \hat\Sigma\right) 
+ 
\left[\lambda(\bar{\sigma}^2-v^2)\bar{\sigma} - c\right]  \frac{1}{V}\int_V d^3x\,\delta\sigma(\mbox{\boldmath$x$}),
\\[1mm]
\Omega^{(2)} &=& \frac{1}{2} \frac{T}{V}
\mathbf{Tr} \, \left( S_0 \hat\Sigma\right)^2 
+ 
\frac{1}{2} \frac{1}{V}\int_V d^3x\,\left[ (\mbox{\boldmath$\nabla$}\delta\sigma(\mbox{\boldmath$x$}))^2  + (\mbox{\boldmath$\nabla$}\delta\pi(\mbox{\boldmath$x$}))^2 \right] 
\nonumber\\
&&
+ \frac{\lambda}{2} (3\bar{\sigma}^2-v^2)  \frac{1}{V}\int_V d^3x\,\left(\delta\sigma(\mbox{\boldmath$x$})\right)^2
+ \frac{\lambda}{2} (\bar{\sigma}^2-v^2)  \frac{1}{V}\int_V d^3x\,\left(\delta\pi(\mbox{\boldmath$x$})\right)^2,
\end{eqnarray}
where
\begin{equation}
S_0^{-1} = i\gamma^\mu \partial_\mu + \mu\gamma^0 - g\bar{\sigma} 
\end{equation}
is the inverse quark propagator Eq.~(\ref{eq:Sinv}) without fluctuations, $S_0$ is its inverse,
and
\begin{equation}
\hat\Sigma = g(\delta\sigma(\mbox{\boldmath$x$})+i\gamma_5 \tau_3 \delta\pi(\mbox{\boldmath$x$}))
\end{equation}
is the quark selfenergy correction due to the fluctuating fields. 

Noting that $S_0$ corresponds to the propagator of a free fermion with mass 
\begin{equation}
   {\bar {M}} = g\bar{\sigma},
\label{eq:Mconst}
\end{equation}   
these expressions are evaluated most easily in momentum space. 
Assuming spatially periodic fields we perform the Fourier expansions
\begin{equation}
       \delta\sigma(\mbox{\boldmath$x$}) = \sum\limits_{\mbox{\scriptsize\boldmath$q$}_k}  \delta\sigma_{\mbox{\scriptsize\boldmath$q$}_k}\, e^{i \mbox{\scriptsize\boldmath$q$}_k \cdot \mbox{\scriptsize\boldmath$x$}},
       \quad
       \delta\pi(\mbox{\boldmath$x$}) = \sum\limits_{\mbox{\scriptsize\boldmath$q$}_k}  \delta\pi_{\mbox{\scriptsize\boldmath$q$}_k}\, e^{i \mbox{\scriptsize\boldmath$q$}_k \cdot \mbox{\scriptsize\boldmath$x$}},
\end{equation}
where $\mbox{\boldmath$q$}_k$ are the elements of the corresponding reciprocal lattice.
Since the meson fields and, thus, their fluctuations are real fields in coordinate space, the Fourier coefficients obey the relations $\delta\sigma_{-\mbox{\scriptsize\boldmath$q$}_k} = \delta\sigma^*_{\mbox{\scriptsize\boldmath$q$}_k}$ and $\delta\pi_{-\mbox{\scriptsize\boldmath$q$}_k} = \delta\pi^*_{\mbox{\scriptsize\boldmath$q$}_k}$.

Taking the infinite-volume limit $V\rightarrow\infty$ one then obtains 
\begin{equation}
\Omega^{(1)} 
= 
\delta\sigma_{\mbox{\scriptsize\boldmath$0$}} \left\{ \lambda(\bar{\sigma}^2-v^2)\bar{\sigma} - c + g^2 \bar{\sigma} F_1 \right\}
\end{equation}
for the linear contribution of the fluctuations to the thermodynamic potential. 
Here we have introduced the loop integral $F_1$, where
\begin{equation}
       F_n= 8 N_c \int \frac{d^3p}{(2\pi)^3}\, T\sum\limits_m \frac{1}{[(i\nu_m +\mu)^2 - {\bf p}^2 -{\bar M}^2]^{n}} 
\end{equation}
with ${\bar M}$ as defined in Eq.~(\ref{eq:Mconst}) and
fermionic Matsubara frequencies $\nu_m = (2m+1)\pi T$.

Note that only the spatially constant ${\mbox{\boldmath$q$}}_k = {\mbox{\boldmath$0$}}$ mode of the fluctuations in the
sigma channel contributes to $\Omega^{(1)}$. 
However, since we have assumed that $\Omega^{(0)}$ corresponds to the lowest homogeneous state, this contribution must
vanish, leading to the gap equation
\begin{equation}
 \lambda(\bar{\sigma}^2-v^2)\bar{\sigma} - c + g^2 \bar{\sigma} F_1 = 0\,.
 \label{eq:gap}
\end{equation} 
Indeed, the same equation can be obtained from the stationary condition $\frac{d\Omega^{(0)}}{d\bar\sigma} = 0$.

Unlike the linear term, the quadratic corrections of the fluctuations to the thermodynamic potential get contributions from all
Fourier modes. One finds
\begin{equation}
\Omega^{(2)} 
= 
-\frac{1}{2}  \sum\limits_{\mbox{\scriptsize\boldmath$q$}_k} 
 \Big\{ 
 |\delta\sigma_{\mbox{\scriptsize\boldmath$q$}_k}|^2 \, D_\sigma^{-1}(q_k)
 + 
 |\delta\pi_{\mbox{\scriptsize\boldmath$q$}_k}|^2 \, D_\pi^{-1}(q_k)
\Big\}\,,
\label{eq:Omega2} 
\end{equation} 
where  $q_k = (0,{\mbox{\boldmath$q$}}_k)$ is the four-momentum vector with vanishing energy and three-momentum
$\mbox{\boldmath$q$}_k$, and
\begin{equation}
D_\mathrm{{\cal M}}^{-1}(q) = q^2 - m_{\mathrm{{\cal M}},\mathrm{t}}^2 + g^2 \Pi_\mathrm{{\cal M}}(q), \quad 
\mathrm{{\cal M}} \in \{\sigma,\pi\},
\end{equation}
are the (unrenormalized) inverse dressed meson propagators at four-momentum $q$, temperature $T$ and chemical potential $\mu$. 
Here
\begin{equation}
m_{\sigma,\mathrm{t}}^2 
= \left. \frac{\partial^2U}{\partial\sigma^2}\right|_{\sigma=\bar\sigma, \pi=0} 
= \lambda(3\bar\sigma^2-v^2)
\quad 
\mbox{and}
\quad
m_{\pi,\mathrm{t}}^2 
= \left. \frac{\partial^2U}{\partial\sigma^2}\right|_{\sigma=\bar\sigma, \pi=0} 
= \lambda(\bar\sigma^2-v^2)
\end{equation}
are the sigma and pion tree-level masses, and $\Pi_\mathrm{{\cal M}}(q)$ denote the corresponding quark-antiquark polarization loops (cf.\ Ref.~\cite{Carignano:2014jla} for further details).
The explicit evaluation yields
\begin{eqnarray}
D_\sigma^{-1}(q) &=& q^2 -2\frac{\lambda}{g^2} {\bar M}^2 - \frac{cg}{{\bar M}} -\frac{1}{2}g^2(q^2-4{\bar M}^2) L_2(q), 
\label{eq:Dsiginv}
\\[1mm]
D_\pi^{-1}(q) &=& q^2 - \frac{cg}{{\bar M}} -\frac{1}{2}g^2 q^2 L_2(q) ,
\label{eq:Dpiinv}
\end{eqnarray} 
where we have used the gap equation (\ref{eq:gap}) to eliminate terms proportional to the loop function $F_1$.
Taking again the infinite-volume limit, the  loop function $L_2$ 
is given by
\begin{eqnarray}
L_2((i\omega_m,{\mbox{\boldmath$q$}}))
=
-8N_c  \int  \frac{d^3p}{(2\pi)^3} \, T\sum\limits_n 
 \frac{1}{[(i\nu_n +i\omega_m+ \mu)^2 - (\mbox{\boldmath$p$}+\mbox{\boldmath$q$})^2 - {\bar M}^2]}
 \nonumber\\
\times \, \frac{1}{[(i\nu_n + \mu)^2 -  \mbox{\boldmath$p$}^2 - {\bar M}^2]} &,
\end{eqnarray}
where $\nu_n$ are again fermionic Matsubara frequencies and $\omega_m$ is a bosonic Matsubara frequency. 
As pointed out above, we only need $L_2$ at zero energy at this point, i.e., $\omega_m = 0$.

In the infinite-volume limit the crystal can take any geometry and size, and therefore the momenta  $\mbox{\boldmath$q$}_k$ of the
reciprocal lattice are not a priori restricted to certain values. 
As can be seen from Eq.~(\ref{eq:Omega2}), the free energy of the homogeneous ground state can thus be lowered by the formation
of inhomogeneous modes if $D_\sigma^{-1}(q) > 0$ or  $D_\pi^{-1}(q) > 0$ in some region of $q = (0,{\mbox{\boldmath$q$}})$.
Note that $q^2 = -{\mbox{\boldmath$q$}}^2$ in this case, so that the inverse propagator of a free meson, 
$D_\mathrm{{\cal M},free}^{-1} = q^2 - m_\mathrm{{\cal M}}^2$ is always negative. The instability is therefore a pure interaction effect, 
as also known, e.g., from P-wave pion condensation in nuclear matter~\cite{Migdal:197x} (see \cite{Kunihiro:1993} for a review). 
In the present model one can distinguish between meson-meson interactions (encoded in the tree-level masses $m_{\mathrm{{\cal M}},\mathrm{t}}$) and quark-meson interactions (encoded in the polarization functions $\Pi_\mathrm{{\cal M}}$).
The latter are identical to the polarization functions in the NJL model, and it was shown in Ref.~\cite{Buballa:2018hux} that they
favor an instability in the sigma channel over an instability in the pion channel.\footnote{The argument, which can be taken over to
the present case to some extent is that $-q^2L_2(q) = {\mbox{\boldmath$q$}}^2L_2(q)$ must be positive to have a chance to create
an instability in the pion channel. But then the corresponding term in the sigma channel, $( {\mbox{\boldmath$q$}}^2+4M^2) L_2(q)$, is even
more positive.}
In the QM model, however, the situation is more complicated because of the tree-level masses. 
At least, if we naively assume the ordering $m_{\sigma,\mathrm{t}}^2 > m_{\pi,\mathrm{t}}^2 > 0$, as for the physical masses in vacuum, we would expect that both masses stabilize the homogeneous phase but less in the pion channel than in the sigma channel.
In order to find out the overall effect we therefore have to evaluate Eqs.~(\ref{eq:Dsiginv}) and (\ref{eq:Dpiinv}) explicitly. 
The resulting stability boundaries of the homogeneous phase will be presented in Sec.~\ref{sec:stabilityRes}.

\subsection{The Real-Kink Crystal ansatz}
\label{sec:rkcForm}
The stability analysis described above has the clear advantage to provide general results 
independent 
 of
the specific shape of the spatial modulation of the order parameter. 
However, since it relies on 
a small-amplitude expansion, it can only 
 provide a sufficient condition for an inhomogeneous phase, while the true inhomogeneous region can be larger.
Thus, in order to complement its results and obtain an estimate of the size of the inhomogeneous window, 
we will also compute the full thermodynamic potential of the QM model for a specific ansatz for the order parameter, the so-called ``real-kink crystal" (RKC).
Aside from the advantage of being a self-consistent ansatz away from the chiral limit \cite{Nickel:2009wj,Schnetz:2005ih}, this RKC
 is also 
 the energetically most favored modulation considered so far in the literature 
 \cite{Buballa:2014tba,Carignano:2012sx,Abuki:2011pf}.

The order parameter is expressed in terms of the Jacobi elliptic functions ${\rm {sn}}, {\rm {cn}}, {\rm {dn}}$, 
\begin{equation}
g\sigma(z) \equiv M(z) = \Delta \Big[ \nu \,{\rm {sn}}(\Delta z, \nu){\rm {sn}}(\Delta z + b, \nu) {\rm {sn}}(b,\nu) + \frac{{\rm {cn}}(b,\nu){\rm {dn}}(b,\nu)}{{\rm {sn}}(b,\nu)} \Big]
\,,
\end{equation}
and is characterized by the three variational parameters $\Delta, \nu$ and $b$, which are determined by minimizing the free energy of the system \cite{Nickel:2009wj}.
For this type of modulation,
an analytical expression for the density of states $\rho(E)$
has been computed, so that the free energy can be obtained without 
having to resort to numerical diagonalization of the inverse quark propagator \cite{Nickel:2009wj,Buballa:2014tba}. One finds 
\begin{align}
\rho(E) &=  \frac{E^2}{\pi^2} \frac{1}{\epsilon} \Big[ \theta(\sqrt{\tilde\nu} - \epsilon) \Big( {\bf E}(\tilde\lambda,\tilde\nu) + C(\nu) {\bf F}(\tilde\lambda,\tilde\nu) \Big) \nonumber\\
&+ \theta(\epsilon - \sqrt{\tilde\nu})\theta(1-\epsilon) \Big( {\bf E}(\tilde\nu) + C(\nu) {\bf K}(\tilde\nu) \Big) \nonumber\\
&+ \theta(\epsilon - 1) \Big( {\bf E}(\lambda,\tilde\nu) + C(\nu) {\bf F}(\lambda,\tilde\nu) + \sqrt{(\epsilon^2 - 1)(\epsilon^2 - \tilde\nu)}/\epsilon \Big) \Big] \,,
\end{align}
where $\epsilon = E/\Delta$, $\tilde\nu = 1-\nu$, $\tilde\lambda = {\rm arcsin}(\epsilon/\sqrt{\tilde\nu})$, $\lambda = {\rm arcsin}(1/\epsilon)$,  $C(\nu) = {\bf E}(\nu)/{\bf K}(\nu) -1$, 
${\bf F}$ and ${\bf K}$ are incomplete and complete elliptic integrals of first kind, respectively, and ${\bf E}$ are the complete and incomplete elliptic integrals of second kind. 

 The thermodynamic potential is then given by
\begin{align}
\Omega  & = - N_f N_c \int_0^\infty dE \, \rho(E) f(\sqrt{E^2 + \delta \Delta^2}) + \frac{1}{2g^2}\ave{(\nabla M)^2} \nonumber\\
 & + \frac{\lambda}{4g^4} \Big[ \ave{M^4} - 2 v^2 g^2 \ave{M^2} + v^4 g^4 \Big] - c \ave{M} \,,
\end{align}
with $\delta = 1/{\rm {sn}}^2(b,\nu) -1$ and 
\begin{equation}
\label{eq:fx}
f(x) = x  + T \log\Big(1+e^{-(x-\mu)/T}\Big)  + T \log\Big(1+e^{-(x+\mu)/T}\Big) \,,
\end{equation}
where the first term corresponds to the vacuum quark contribution.

The meson potential depends on the following spatial averages of the order parameter over a period: 
\begin{align}
\ave{M} & = \Delta \Big[ Z(b,\nu) + \sqrt{\frac{\delta(1+\delta-\nu)}{1+\delta}} \Big] \,, \\
\ave{M^2} & = \Delta^2 \Big[ \delta - \nu - 2 C(\nu) \Big] \,, \\
\ave{M^4} & =  \Delta^4 \Big[ (\delta - \nu)^2 - \frac{8}{3}\Big[\nu + C(\nu) (2 + 3 \delta - \nu)\Big]  \nonumber\\ &- 
 4 \sqrt{\delta(1 + \delta)(1+\delta-\nu)} Z(b,\nu) \Big] \,, \\
\ave{(\nabla M)^2} & = \Delta^4 \frac{4}{3} \Big[ \nu + (2-\nu+3\delta) C(\nu) + 3 \sqrt{\delta(1 + \delta)(1+\delta-\nu)}  Z(b,\nu) \Big] \,,
\end{align}
where $Z$ is the Jacobi Zeta function. 

Before getting to our results for the model phase structure, let us now discuss how the model parameters are fixed.

\section{Parameter fixing}
\label{sec:parameters}

As standard, we determine the model parameters by fitting masses and the pion decay constant in vacuum. 
Thereby, in order to systematically investigate the effect of the explicit chiral-symmetry breaking, we first set the coupling $c$ 
equal to  zero and fix the remaining parameters $g$, $\lambda$ and $v^2$ in the chiral limit. After that, we consider $c\neq 0$
but keep the other parameters at their chiral-limit values.

For fixing $g$, $\lambda$ and $v^2$ in the chiral limit we follow Ref.~\cite{Carignano:2014jla}, where this was done by fitting the vacuum values of the pion decay constant $f_\pi$, of the sigma-meson mass $m_\sigma$, and of the constituent quark mass. For homogeneous matter we can identify the latter with $\bar M$ as defined in
 Eq.~(\ref{eq:Mconst}) with $\bar\sigma$ being the homogeneous sigma field which minimizes $\Omega^{(0)}$. In vacuum, i.e., at $T=\mu=0$, we expect that it also minimizes $\Omega_\mathrm{MFA}$, since phenomenologically the vacuum is homogeneous. This turns out to be true in our model as well, at least up to quadratic-order fluctuations. 
For $m_\sigma$ and $f_\pi$ it was shown in Ref.~\cite{Carignano:2016jnw} that it is crucial to fit the pole mass 
and to take into account the renormalization of the pion wave function, 
corresponding to the pole of $D_\sigma$ and the residue of $D_\pi$, respectively. 
The resulting expressions are (see Refs.~\cite{Carignano:2014jla,Carignano:2016jnw} for details)
\begin{eqnarray}
       g^2 &=& \frac{{\bar M}^2_0}{f_{\pi,0}^2 + \frac{1}{2}{\bar M}_0^2 L_2^{(\mathrm{vac},0)}(0)}  \,,
\label{eq:g2RP}
\\
      \lambda
       &=&
       2g^2 \frac{m_{\sigma,0}^2}{4{\bar M}_0^2} 
       \left[ 1
       - \frac{1}{2}g^2
       \left(1 - \frac{4{\bar M}_0^2}{m_{\sigma,0}^2} \right) L_2^{(\mathrm{vac},0)}(m_{\sigma,0}) 
       \right]\,,
\label{eq:lambdaRP}
\\
       v^2 &=& \frac{{\bar M}_0^2}{g^2} + \frac{g^2 F_1^{(\mathrm{vac},0)}}{\lambda}  \,,
       \label{eq:v2}
\end{eqnarray}
where the subscript $0$ in ${\bar M}_0$, $m_{\sigma,0}$ and $f_{\pi,0}$ indicates that these quantities correspond to the
vacuum values in the chiral limit. Likewise $F_1^{(\mathrm{vac},0)}$ and $L_2^{(\mathrm{vac},0)}$ are the loop integrals $F_1$ and
$L_2$ evaluated in vacuum and with ${\bar M}={\bar M}_0$. Moreover,  $L_2^{(\mathrm{vac},0)}(m_{\sigma,0})$ means that the function 
$L_2((i\omega_m,{\mbox{\boldmath$q$}}))$ is analytically continued to the real time-like momentum 
$q=(m_{\sigma,0},\mbox{\boldmath$0$})$. The explicit expressions can be found in  Ref.~\cite{Carignano:2016jnw}. 

Having fixed $g$, $\lambda$ and $v^2$ in this way, we turn on the chiral-symmetry breaking term by varying the parameter $c$. 
The most important consequence is that the pion, which is massless in the chiral limit in agreement with the Goldstone theorem, gets a non-vanishing mass. We can therefore relate the parameter $c$ to the pion pole mass, implicitly given by $D_\pi^{-1}(q=(m_{\pi},\mbox{\boldmath$0$}))=0$. We then get from Eq.~(\ref{eq:Dpiinv})
\begin{equation}
       cg
       =
       m_\pi^2 {\bar M} 
       \left[ 1
       - \frac{1}{2}g^2 L_2^{(\mathrm{vac})}(m_{\pi}) 
       \right]\,,
\label{eq:cg}
\end{equation}
where $L_2^{(\mathrm{vac})}(m_{\pi})$ is the function $L_2((i\omega_m,{\mbox{\boldmath$q$}}))$ evaluated in vacuum and analytically continued to the real time-like momentum $q=(m_{\pi},\mbox{\boldmath$0$})$.
Note, however, that the quark mass $\bar M$, which also enters the function $L_2^{(\mathrm{vac})}$, is {\it not} the chiral-limit value
${\bar M}_0$, as in Eqs.~(\ref{eq:g2RP}) -- (\ref{eq:v2}), but related to the solution of the gap equation (\ref{eq:gap}), including the constant $c$. For a fixed value of $m_\pi$, Eq.~(\ref{eq:cg}) must therefore be solved self-consistently together with Eq.~(\ref{eq:gap}).

Finally, we note that the vacuum parts of loop integrals $F_1$ and $L_2$, as well as their chiral-limit versions, are
ultraviolet divergent and must be regularized in order to get meaningful results.\footnote{{In earlier QM-model studies
the divergent vacuum parts have often been dropped completely, arguing that their effects can be absorbed in the model parameters~\cite{Scavenius:2000qd,Nickel:2009wj}. As shown however in Ref.~\cite{Skokov:2010sf}, this so-called `standard mean-field approximation' causes artifacts in the phase diagram. Therefore we take into account the vacuum contributions of the quark loops explicitly. }
}
Again following Refs.~\cite{Carignano:2014jla,Carignano:2016jnw}, we use Pauli-Villars regularization with three regulators, controlled by the cutoff parameter $\Lambda$.
As a consequence,  the model parameters for fixed values of ${\bar M}_0$,  $m_{\sigma,0}$, $f_{\pi,0}$, and $m_\pi$ depend on $\Lambda$. 

\begin{figure}
\resizebox{0.95\columnwidth}{!}{%
  \includegraphics{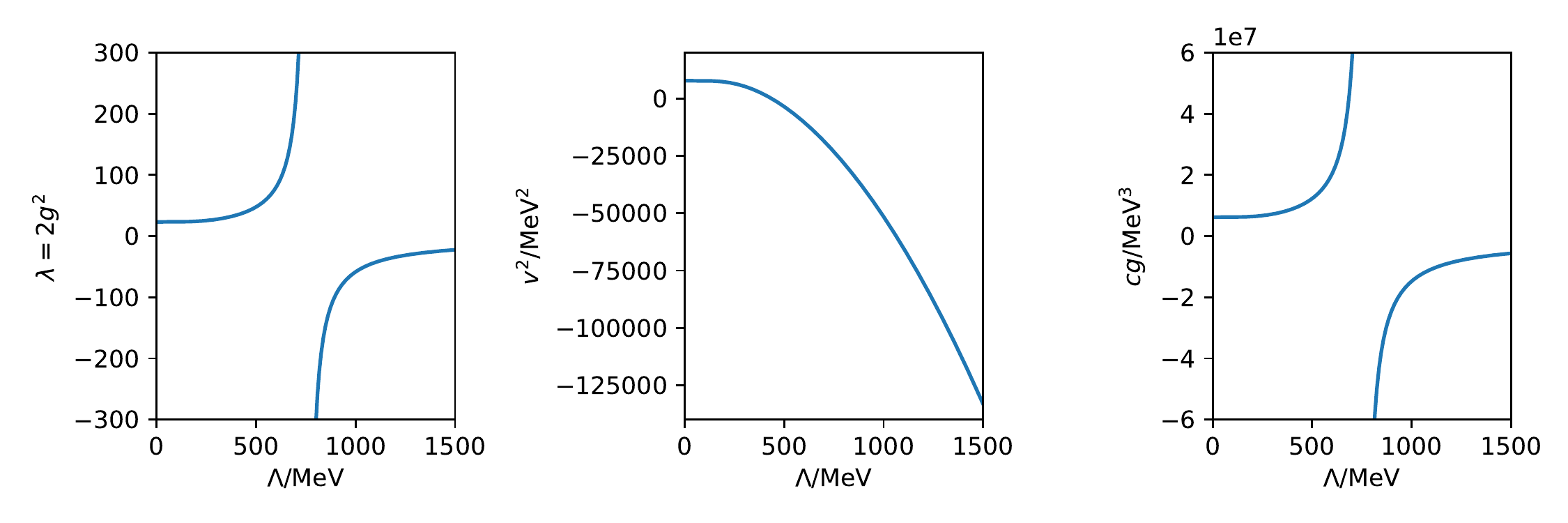} }
\caption{Model parameters as functions of the Pauli-Villars cutoff parameter $\Lambda$.
The parameter $c$ has been fitted to a vacuum pion mass $m_\pi = 140$~MeV. 
}
\label{fig:const}       
\end{figure}

In the following, we will always fix our model in the chiral limit by choosing ${\bar M}_0=300$~MeV,  $m_{\sigma,0}=600$~MeV, and $f_{\pi,0}=88$~MeV. In particular we have $m_{\sigma,0}=2{\bar M}_0$, in which case Eq.~(\ref{eq:lambdaRP}) simplifies to $\lambda = 2g^2$.
The corresponding values of $\lambda$ and $v^2$ as functions of $\Lambda$ are displayed in the first two
panels of Fig.~\ref{fig:const}. The results agree with those in Refs.~\cite{Carignano:2014jla,Carignano:2016jnw}, where the same vacuum observables have been fitted. In addition, we show in Fig.~\ref{fig:const} the parameter $c$, multiplied with $g$, for 
$m_\pi = 140$~MeV. 

\begin{figure}
\resizebox{0.95\columnwidth}{!}{%
  \includegraphics{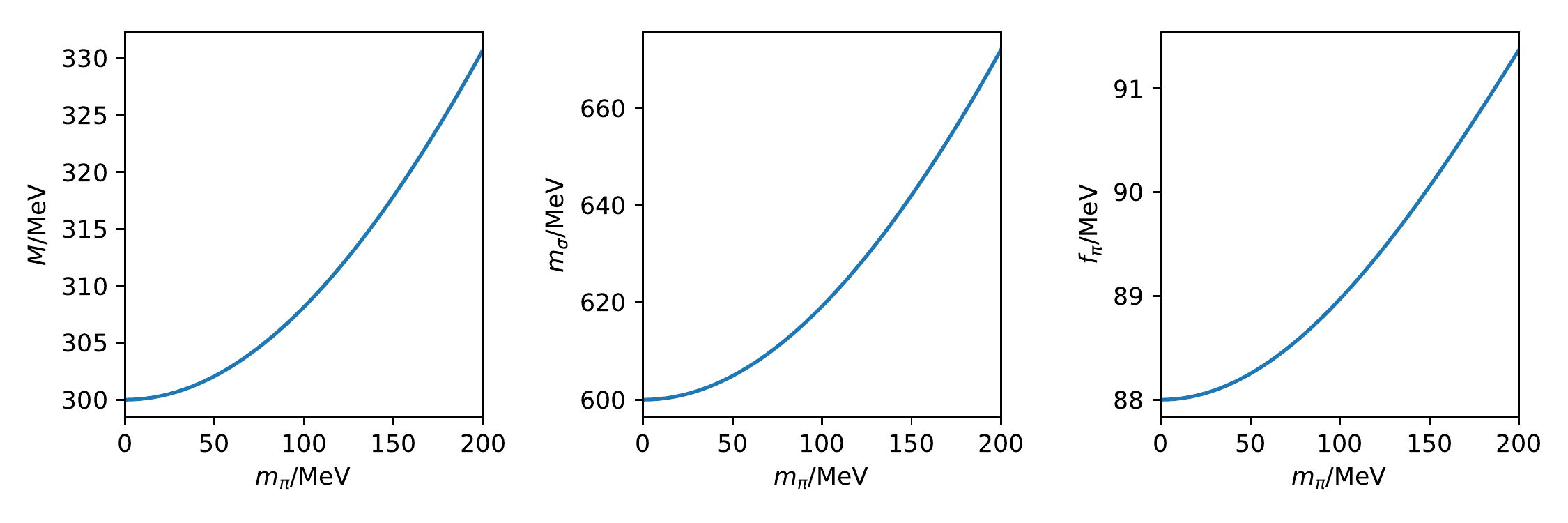} }
\caption{Vacuum properties as functions of the pion mass in the renormalized limit ($\Lambda = 5$~GeV):
constituent quark mass (left), sigma-meson mass (center), and pion decay constant in
the Goldberger-Treiman approximation, Eq.~(\ref{eq:fpiGT}) (right).
}
\label{fig:obs}       
\end{figure}

Since the QM model is renormalizable, all observables should remain finite in the limit $\Lambda \rightarrow \infty$. 
As demonstrated in Ref.~\cite{Carignano:2014jla}, this is also true for the phase diagram. It was found that the results remain practically unchanged when $\Lambda$ exceeds $2$~GeV, so that in practice  $\Lambda=5$~GeV can be considered as the 
``renormalized limit''.\footnote{A formal one-loop renormalization of the model within dimensional regularization, including an application to inhomogeneous phases, has been performed in Ref.~\cite{Adhikari:2017ydi}.}
In Fig.~\ref{fig:obs} we show how the vacuum values of $M$, $m_\sigma$ and $f_\pi$ vary as functions of $m_\pi$ in the renormalized limit. 
By construction, they take of course their fit values in the chiral limit, i.e., at $m_\pi=0$. 
With increasing $m_\pi$ they increase as well but stay finite, even for arbitrarily large values of $\Lambda$.
We note that the value of $f_\pi$ for the physical pion mass $m_\pi \sim 140$~MeV is too small compared with the empirical value of 92.2~MeV~\cite{PDG}.
This could be cured by slightly changing the fit values in the chiral limit (which are admittedly somewhat ad-hoc) but it is not our intention here to perform a precision fit. Moreover, in Fig.~\ref{fig:obs}, $f_\pi$ has been calculated as~\cite{Carignano:2014jla}
\begin{equation}
       f_\pi^2 = \frac{\bar M^2}{Z_\pi g^2} =  \frac{\bar M^2}{g^2} \left(1 -\frac{1}{2} g^2 L_2^{(\mathrm{vac})}(m_\pi)\right),      
\label{eq:fpiGT}          
\end{equation}
which corresponds to the quark-level Goldberger-Treiman relation and is strictly speaking only valid in the chiral limit (cf.~Eq.~(\ref{eq:g2RP})). 

A more severe problem is that  $g^2$ and $\lambda$ diverge at 
the point when $L_2^{(\mathrm{vac},0)} = -2f_{\pi,0}^2/{\bar M}_0^2$.
Within our regularization scheme and for our parameters  this happens at $\Lambda = \Lambda^* \approx 757$~MeV.
Beyond this point, $g^2$ and $\lambda$ even turn negative, see Fig.~\ref{fig:const}, and,
related to this, $\Omega_\mathrm{MFA}$ is no longer bounded from below in this regime~\cite{Carignano:2014jla,Skokov:2010sf}.
Moreover, a negative $g^2$ obviously means that the Yukawa coupling $g$ is imaginary. Hence, forcing the constituent quark mass, Eq.~(\ref{eq:Mconst}), to stay real, the field expectation value $\bar\sigma$ becomes imaginary as well, in contradiction to our original assumption of $\sigma$ and $\pi$ being real fields. 
Although it has been argued in Ref.~\cite{Skokov:2010sf} that the unbounded potential is a known one-loop artifact and should be cured at higher orders, this is clearly worrisome.
On the other hand, the phase diagram changes smoothly when passing through $\Lambda = \Lambda^*$,
i.e., focusing only on the phase diagram, one would not even notice that the problem exists.
In Sec.~\ref{sec:phase} we will therefore discuss results for the renormalized limit, ignoring the inconsistencies, as well as for $\Lambda = \Lambda^*$, being the largest possible cutoff outside the problematic regime.\footnote{Incidentally, we note that, if we restrict ourselves to the chiral limit and the case $m_\sigma=2{\bar M}_0$, in the limit $\Lambda = \Lambda^*$ the meson potential reduces to $U(M^2 = \sigma^2 + \pi^2) =  - M^2 v^2$ (up to an infinite constant), and thus {the model}
becomes equivalent to the NJL model fitted to the same vacuum quantities
upon the identification  $v^2(\Lambda^*) = -1/(4G_\mathrm{NJL})$, $G_\mathrm{NJL}$ being the coupling constant of the four-fermi interaction in the NJL model. 
}

\section{Phase structure}
\label{sec:phase}

We are now ready to discuss our results for the model phase structure, starting from the stability analysis to determine the boundary where inhomogeneous phases
become favored.

\subsection{Stability analysis} 
\label{sec:stabilityRes}

\begin{figure}
\resizebox{0.49\columnwidth}{!}{%
  \includegraphics{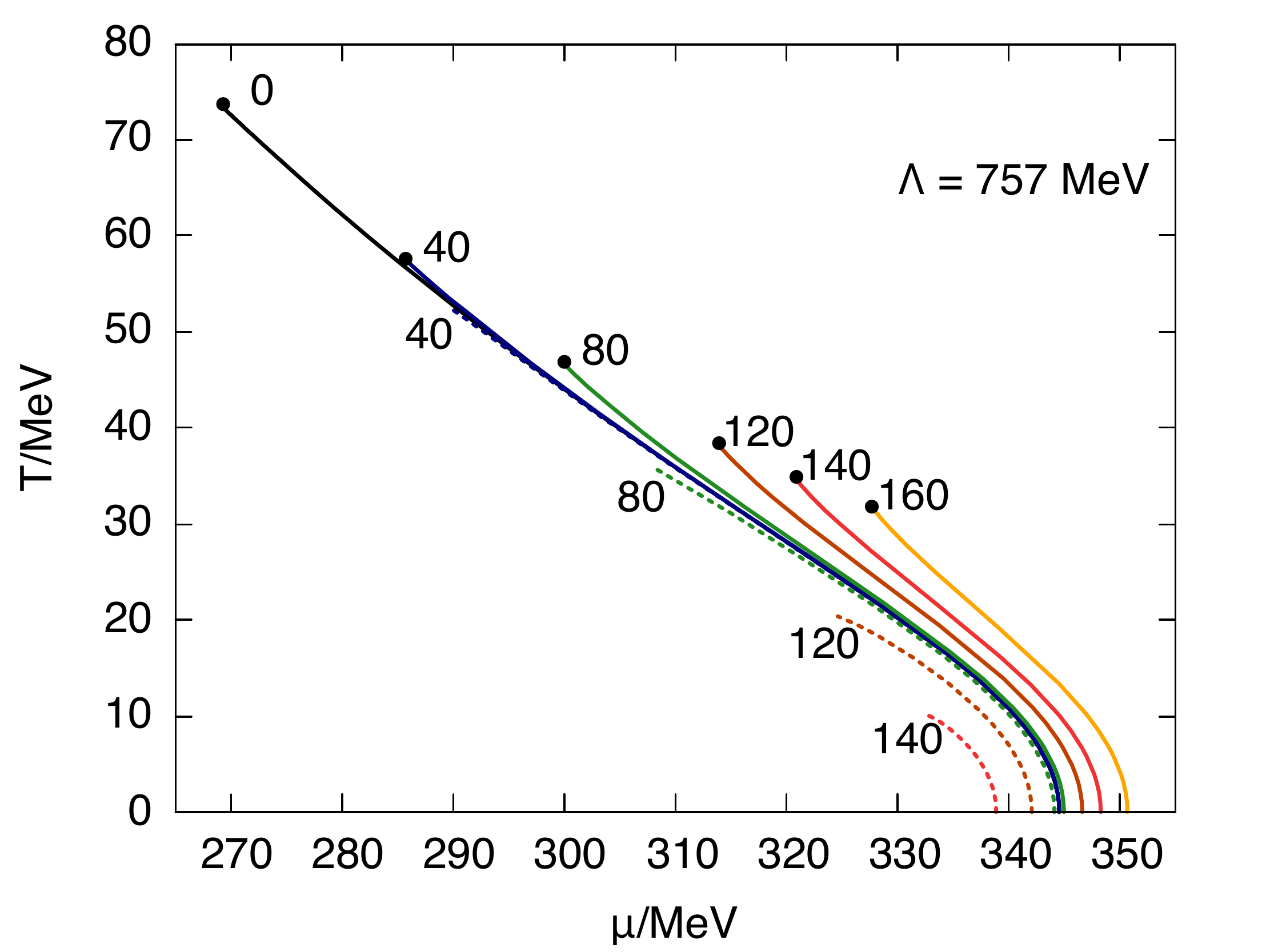} }
\resizebox{0.49\columnwidth}{!}{%
  \includegraphics{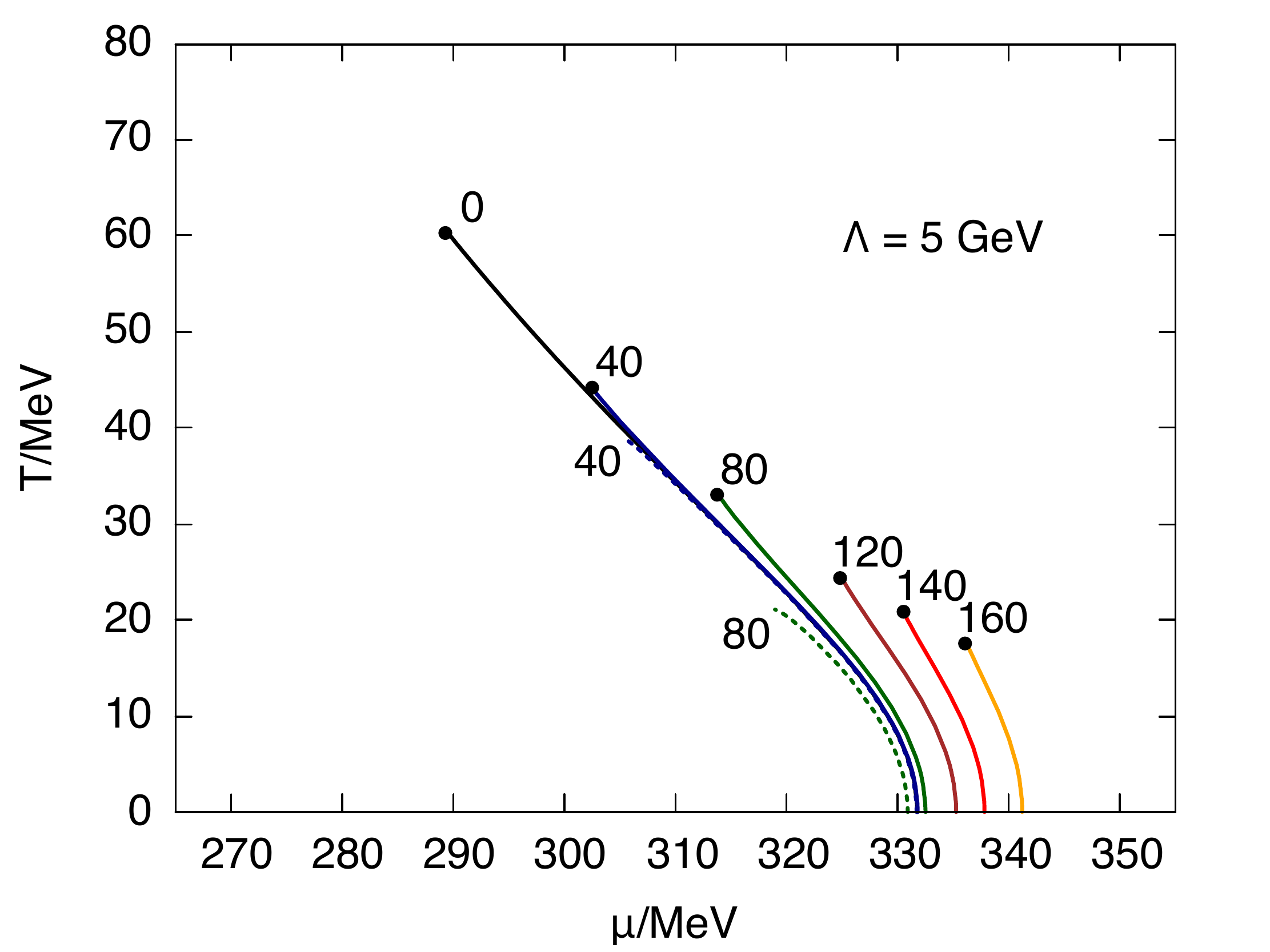} }
\caption{Stability boundaries of the homogeneous regions with respect to scalar (solid lines) or pseudoscalar (dashed lines) inhomogeneous fluctuations for $\Lambda =\Lambda^*= 757$~MeV (left panel) and for the ``renormalized case'' ($\Lambda = 5$~GeV, right panel). The different lines correspond to different values of the pion mass, in MeV as indicated by the labels.
The unstable regions lie to the left of these lines, i.e., the solid lines thus correspond to the right (upper $\mu$) boundaries of the
inhomogeneous phase. Note that the left (lower $\mu$) boundaries, which cannot be determined by the stability analysis, are not shown.
\label{fig:grenze}   }    
\end{figure}

In Fig.~\ref{fig:grenze} we show the stability boundaries of the homogeneous phase with respect to inhomogeneous fluctuations.
 More precisely, we show the lines where $D_\mathcal{M}^{-1}(q)$ just touches the zero-line at some value of $q = (0,{\mbox{\boldmath$q$}}\neq{\mbox{\boldmath$0$}})$, both for $\mathcal{M}=\sigma$ or $\mathcal{M}=\pi$,
for different values of the PV regulator and the vacuum pion mass.
We recall that this type of analysis relies on the assumption that the spatially modulated order parameters
are small, and thus can only give reliable results for second-order phase boundaries. According to explicit calculations with
certain modulations, this is typically the case at the right phase boundary of the inhomogeneous region, while the left boundary 
cannot reliably be determined by the stability analysis. We will confirm this below in Sec.~\ref{sec:pdfull}.

As demonstrated in Ref.~\cite{Carignano:2014jla} for the chiral limit, incorporating 
 vacuum fluctuations shrinks the size of the inhomogeneous phase,
which nevertheless survives in the renormalized limit.
As we move away from the chiral limit, the stability lines in the two channels split, with the sigma line becoming the only relevant one since it is the first to appear when coming from the stable homogeneous region at higher chemical potential. Moreover, the pion lines decrease rapidly with growing $m_\pi$ and eventually disappear from the phase diagram. On the other hand, albeit reduced, the instability in the sigma channel is still present for a physical pion mass in the renormalized limit, so that we still expect an inhomogeneous phase driven by the scalar condensate.

\begin{figure}
\begin{center}
\resizebox{0.4\columnwidth}{!}{%
  \includegraphics{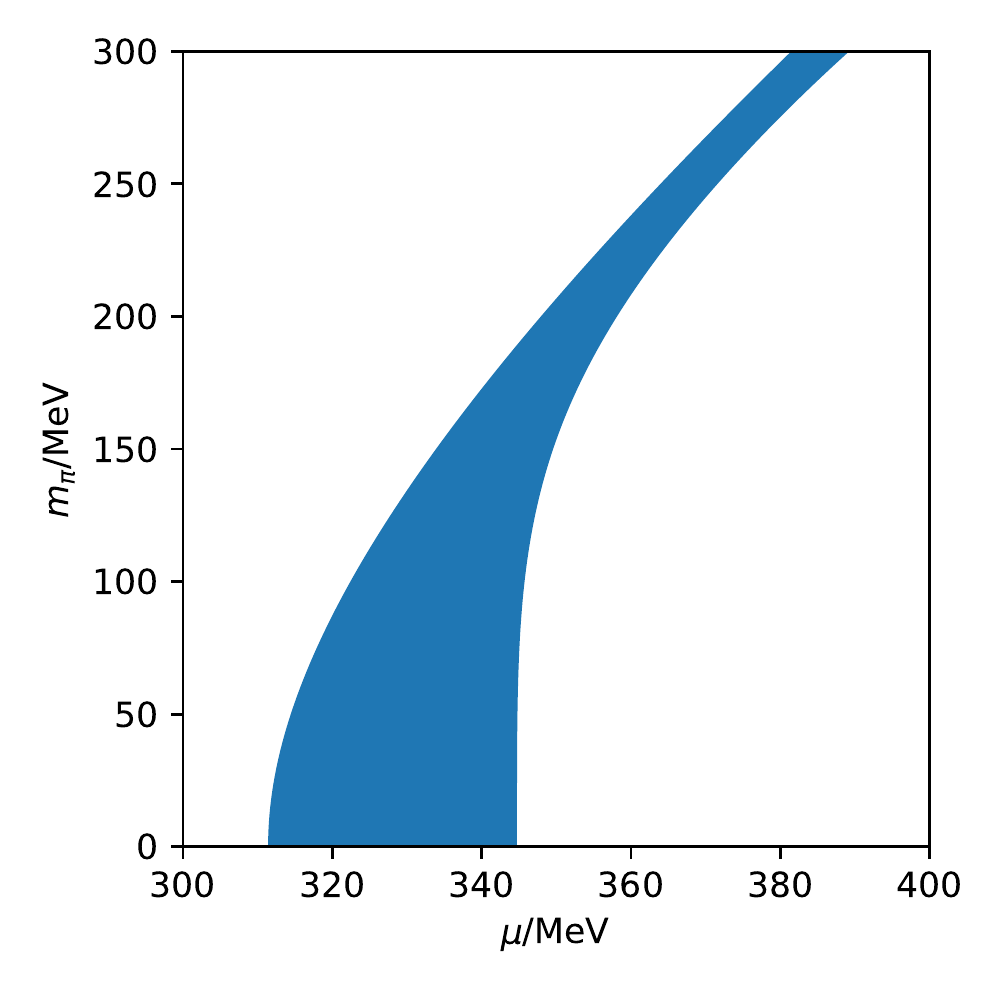} }
\resizebox{0.4\columnwidth}{!}{%
  \includegraphics{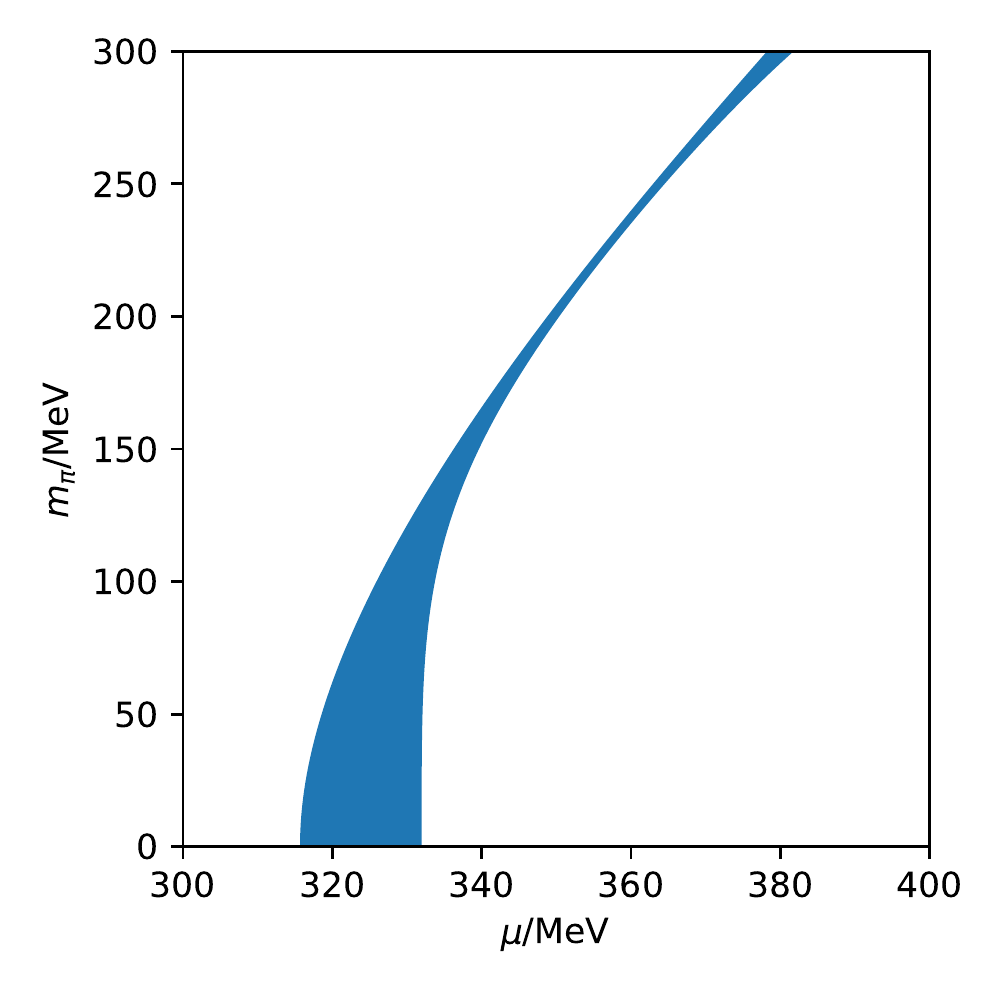} }
  \end{center}
\caption{Instability regions in the $\mu-m_\pi$ plane for  $\Lambda=757$ MeV (left) and the renormalized limit (right). 
\label{fig:grenzempi}   }    
\end{figure}

In Fig.~\ref{fig:grenzempi} we show the extension of the whole instability region, i.e., the whole chemical-potential interval where $D_\sigma^{-1}(q)$ is positive for some $q = (0,{\mbox{\boldmath$q$}}\neq{\mbox{\boldmath$0$}})$, 
at vanishing temperature and varying pion mass. We find that
even in the renormalized limit a finite window of instability persists for all values of $m_\pi$ considered. While going from the chiral limit to a physical pion mass 
reduces the size of the instability region, when $m_\pi$ becomes very large its extension starts increasing again, a similar behavior to the one observed in the NJL model away 
from the chiral limit \cite{Buballa:2018hux}.\footnote{At even larger pion masses the instability region joins with the ``inhomogeneous continent'', which in the chiral limit appears at high chemical potentials (see Refs.~\cite{Carignano:2014jla,Carignano:2011gr}).}

It is worth emphasizing that the outcome of our stability analysis is not in discrepancy with the renormalized-limit results of 
Ref.~\cite{Andersen:2018osr}, 
where it was found for a CDW ansatz that the inhomogeneous phase becomes disfavored against homogeneous solutions
already at $m_\pi =37$~MeV: This is due to the fact that the CDW ansatz enforces equal amplitudes for the scalar and pseudoscalar channels, the latter being disfavored 
according to our stability analysis. 
A different ansatz which allows for inhomogeneous condensation only in the $\sigma$ channel on the other hand should be thermodynamically favored over 
homogeneous matter in this region of the phase diagram.
 In the following section we will demonstrate this with the specific example of the RKC modulation.

\subsection{Full phase diagram for the RKC ansatz}
\label{sec:pdfull}

Having determined the behavior of the instability lines for a generic inhomogeneous order parameter away from the chiral limit, 
we now  compute the full phase diagram for a specific ansatz, the RKC modulation introduced in Sec.~\ref{sec:rkcForm}.
 To be consistent with the stability analysis, we regularize
the vacuum contribution of Eq.~(\ref{eq:fx}) using three Pauli-Villars counterterms \cite{Nickel:2009wj}. 

\begin{figure}
\resizebox{0.49\columnwidth}{!}{%
  \includegraphics{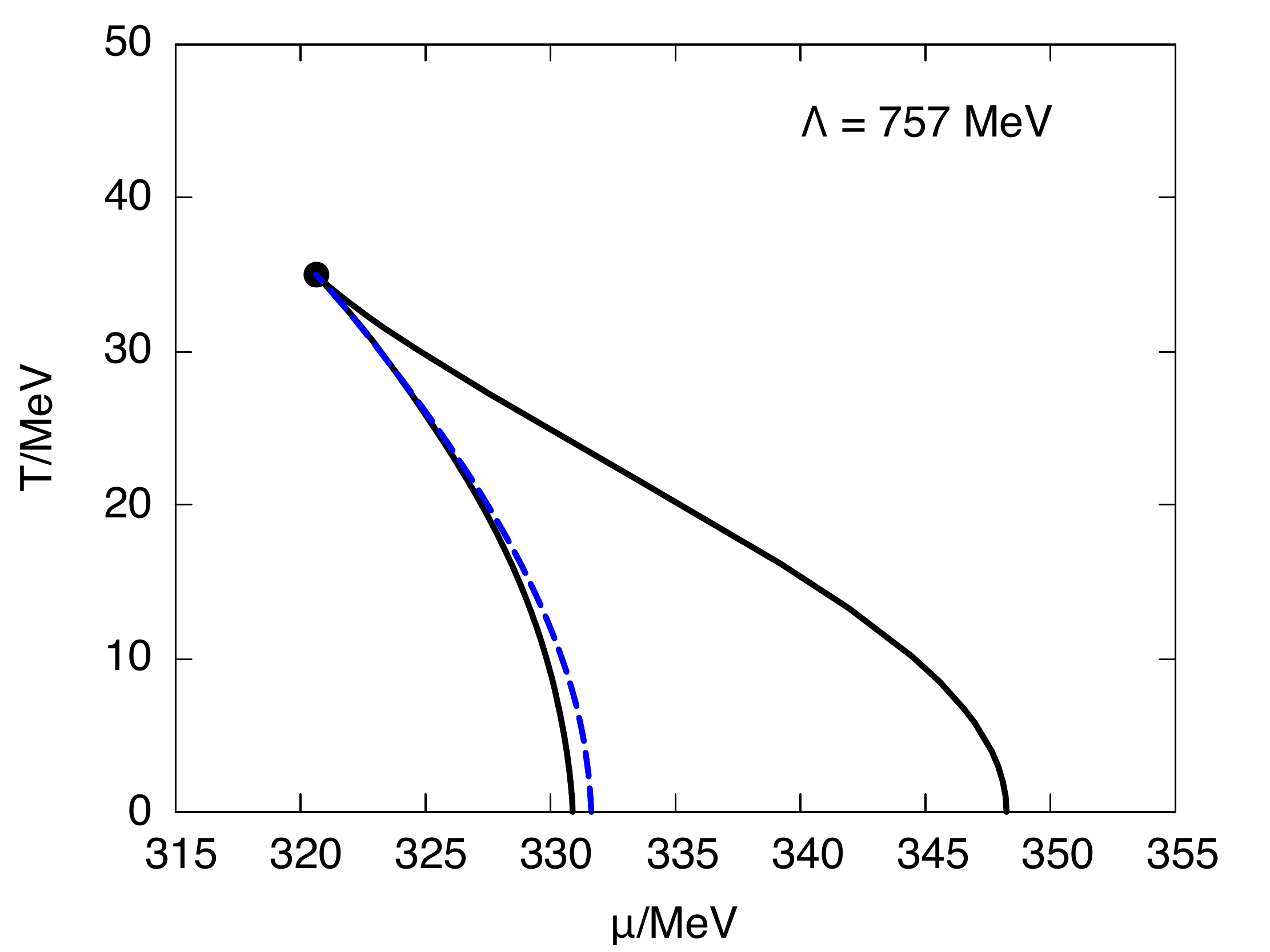} }
  \resizebox{0.49\columnwidth}{!}{%
  \includegraphics{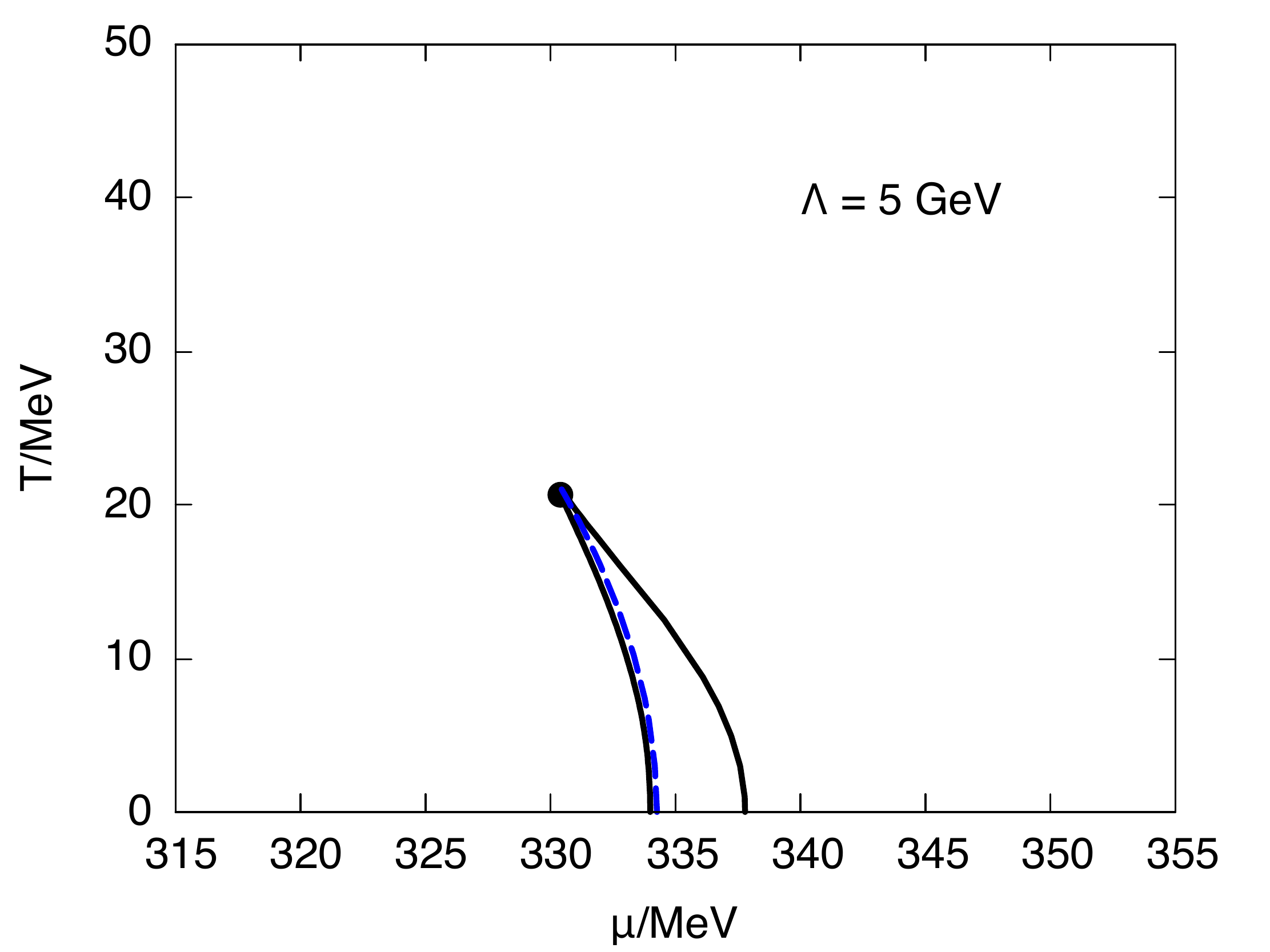} }
\caption{Full phase diagram for $\Lambda=757$~MeV (left) and the renormalized limit (right) for a physical vacuum pion mass $m_\pi = 140$ MeV. The tip of the inhomogeneous phase coincides with the position
of the CEP for homogeneous order parameters. Solid black lines denote the boundaries of the inhomogeneous phase
for the RKC ansatz, while the blue dashed lines are the left edges of the instability region found with the 
stability analysis. 
For the right boundary of the inhomogeneous phase both approaches yield coinciding results.
\label{fig:full1}   }    
\end{figure}

In Fig. \ref{fig:full1} we show the phase diagram for a physical pion mass $m_\pi = 140$ MeV, both for  $\Lambda = 757$ MeV and in the renormalized limit. As expected, we find an inhomogeneous phase
whose right boundary coincides with the instability line for the sigma channel found in the previous section. For comparison, we show in the figure also the left edge of the instability region, which, as expected, falls inside the inhomogeneous phase. 
In fact, the left edge of the instability region coincides with the first-order phase boundary one finds when the model is restricted to homogeneous phases, and just reflects the discontinuous change of the expansion point.
In other words: While the chirally almost restored phase just to the right of the first-order boundary is unstable against small inhomogeneous fluctuations, the larger homogeneous condensates to the left make it at least metastable.
Our results with the RKC ansatz show however, that it is still possible to lower the free energy of the system by large
inhomogeneous fluctuations in this region.

Furthermore, our numerical results suggest that
the tip of the inhomogeneous phase, 
the so-called pseudo-Lifshitz point (PLP),\footnote{ A Lifshitz point can be defined as the point where three different phases (chirally broken, restored and the spatially inhomogeneous one) meet, so in this case we should be referring to it as a pseudo-Lifshitz point, since away from the chiral limit there is only a crossover above the CEP.
}
coincides with the location of the 
 CEP
obtained when restricting the analysis to homogeneous matter. 
This is similar to what happens in the NJL model, and it can be understood in a general way via a Ginzburg-Landau analysis, as discussed in the following section.

\subsection{Ginzburg-Landau expansion} 

The Ginzburg-Landau (GL) expansion is a systematic expansion of the thermodynamic potential in powers of the order parameter and its gradients. It is a powerful tool which
allows to determine precisely the locations of the CEP and the PLP where both the amplitude and the gradients of the spatially modulated order parameter approach zero.
In the following we want to use this method, which has been applied to the NJL model in Ref.~\cite{Buballa:2018hux},
to study the behavior of the CEP and PLP in the QM model away from the chiral limit.

For this,
following the steps performed in Ref.~\cite{Buballa:2018hux} for the NJL model, neglecting pseudoscalar fluctuations we write
 again $g\sigma({\bf x}) = {\bar M} + \delta M({\bf x})$ and get to $ \Omega[M] = \Omega[{\bar M}]  + \frac{1}{V} \int d^3x\, \delta\omega[\bar M,\delta M({\bf x})]$ with 
\begin{equation}
      \delta\omega =   \alpha_1 \delta M
        + \alpha_2 \delta M^2 + \alpha_3 \delta M^3
       + \alpha_{4,a} \delta M^4   + \alpha_{4,b} (\nabla \delta M)^2
       + \dots     \,,
\label{eq:Omega_GL_m}
\end{equation}
where the GL coefficients $\alpha_i$ depend on $T$, $\mu$ and $\bar M$.
As shown in Ref.~\cite{Buballa:2018hux},
 we can localize the CEP as the point where the GL coefficients  $\alpha_1 = \alpha_2 = \alpha_3 = 0$  
 (the condition $\alpha_1=0$ simply enforces the gap equation for the background ${\bar M}$), whereas the  PLP  is identified as
the point where both the quadratic and the first non-vanishing gradient term become zero: 
$\alpha_1 = \alpha_2 = \alpha_{4,b} = 0$. 
For the relevant coefficients we find
\begin{align}
\alpha_1 & = \frac{\lambda}{g^4}{\bar M}({\bar M}^2 - v^2g^2) - \frac{c}{g} + {\bar M} F_1 \,,
\\
\alpha_2 & =  \frac{\lambda}{2g^4}  (3{\bar M}^2 - g^2 v^2) + \frac{1}{2} F_1 +  {\bar M}^2 F_2 \,,
\\
\alpha_3 & =  4\bar M \left ( \frac{\lambda }{4g^4}  + \frac{1}{4}  F_2 +  \frac{1}{3} {\bar M}^2 F_3 \right) \,,
\\
\alpha_{4,b} & = \frac{1}{2g^2} +  \frac{1}{4} F_2 +  \frac{1}{3}  {\bar M}^2 F_3 \,.
\label{eq:alpha4bres}
\end{align}

Upon close inspection, we see that when $m_{\sigma,0} = 2{\bar M}_0$, and thus $\lambda = 2g^2 $, the coefficients $\alpha_3$ and $ \alpha_{4,b}$ are proportional to each other, like in the NJL model, 
and as a result, the CEP and the PLP coincide, supporting the numerical results of our previous section.
In Fig.~\ref{fig:grenze} we have indicated the positions of these points for various values of $m_\pi$ by black dots.

\section{Conclusions}
\label{sec:conclusions}

We have investigated inhomogeneous phases in the renormalized limit of the quark-meson model away from the chiral limit.
Both the effect of the vacuum quark fluctuations in the QM model~\cite{Carignano:2014jla}
as well as the inclusion of an explicit chiral-symmetry breaking term~\cite{Buballa:2018hux} 
are known to shrink the size of the inhomogeneous window
in the phase diagram, so it is natural to ask whether an inhomogeneous phase survives at all when both effects are taken into account.
A first investigation in this direction found that if  one restricts the analysis to a CDW modulation the inhomogeneous phase quickly disappears and is not present for physical pion masses~\cite{Andersen:2018osr}. On the other hand,
it is known that other types of spatial modulations of the order parameter are usually thermodynamically more favored. 

Thus, in order to obtain a modulation-agnostic answer, we first looked for the appearance of instabilities of homogeneous matter towards inhomogeneous phases for arbitrary shapes of the order parameter, 
and found that 
such an instability exists even for pion masses above the physical one.
 This instability occurs with respect to scalar fluctuations, 
whereas the instability in the pseudoscalar channel disappears quickly as $m_\pi$ increases.
 Explicit chiral-symmetry breaking thus strongly suppresses 
fluctuations in the pseudoscalar channel, explaining the rapid disappearance of the CDW modulation from the phase diagram. 

We supported these findings with an explicit calculation of the model phase diagram considering a specific modulation of the order parameter involving only the scalar channel, the so-called real-kink crystal, which provides a 
self-consistent ansatz away from the chiral limit, and checked that indeed the inhomogeneous phase has a non-vanishing extension for a physical $m_\pi$ in the renormalized limit of the model.

The presence of inhomogeneous phases thus seems to be a robust model feature, even though the size of the inhomogeneous window found in this work is relatively small.
In particular, our GL analysis revealed that for $m_{\sigma,0} = 2{\bar M}_0$ the PLP, i.e., the tip of the inhomogeneous phase, 
coincides with the CEP of the first-order phase boundary in the homogeneous case and, hence, the inhomogeneous phase is as
robust a feature of the model as the existence of a first-order phase transition if the analysis is restricted to homogeneous phases.
We must keep in mind, however, that the present analysis has been performed in mean-field approximation.
It is thus an interesting question, both in the chiral limit and away from it, whether these findings remain valid if fluctuation effects
are taken into account. 
Investigations of such questions are presently subject
of intese research~\cite{Pisarski:2018bct,Pisarski:2020dnx}, particularly within the functional renormalization-group approach~\cite{Fu:2019hdw,Tripolt:2017zgc,SBS} 
or by performing lattice simulations for lower-dimensional models~\cite{Pannullo:2019bfn,Lenz:2020bxk}.

\subsection*{Acknowledgments}	
We thank B.-J.~Schaefer, M.J.\ Steil, and M.~Winstel for useful comments on the manuscript,
and the anonymous referee for spotting a mistake in Fig.~3 of the first version. 
M.B.\ and L.K.\ acknowledge support by the Deutsche Forschungsgemeinschaft (DFG, German Research Foundation) through the CRC-TR 211 ``Strong-interaction matter under extreme conditions" - project number 315477589 - TRR 211.
S.C. has been supported by the projects FPA2016-81114-P and FPA2016-76005-C2-1-P (Spain), and by the project 2017-SGR-929 (Catalonia).

\end{document}